\documentclass[aps,pra,reprint,showpacs]{revtex4-2}
\usepackage{amsfonts}
\usepackage{amsmath}
\usepackage{amssymb}
\usepackage{charter}
\usepackage{graphicx}

\begin{document}

\title{Information Geometry of Four-Parameter Single-Qutrit States: From
Quantum to Semiclassical Geometric Tensors}
\author{Xue-xiang Xu$^{1,\dag }$, Kai-xu Cai$^{1}$, Xue-feng Zhan$^{2}$,
Hong-chun Yuan$^{3,\ddag }$ }
\affiliation{$^{1}$School of Physics, Jiangxi Normal University, Nanchang 330022,
People's Republic of China;\\
$^{2}$School of Microelectronics, Jiangxi Normal University, Nanchang
330022, People's Republic of China;\\
$^{3}$School of Electrical and Information Engineering, Changzhou Institute
of Technology, Changzhou 213032, People's Republic of China\\
$^{\dag }$xuxuexiang@jxnu.edu.cn\\
$^{\ddag }$yuanhc@czu.cn}

\begin{abstract}
The gap between the classical and quantum Fisher information matrices (CFIM
and QFIM) separates what is operationally accessible through measurements
from what is intrinsic to a quantum state. In the multiparameter setting,
this quantum obstruction is generically not saturable. Motivated by the
recently introduced semiclassical geometric tensor (SCGT), we perform an
explicit information-geometric study for a class of pure four-parameter
single-qutrit states (FPSQSs). These states are probed by a one-parameter
family of measurements that interpolates between an uninformative POVM and a
sharp projective measurement. We derive closed-form expressions for the
CFIM, the quantum geometric tensor (QGT), and the SCGT. We show that the
SCGT reproduces the QGT in the projective limit and vanishes in the trivial
limit. The real part of the SCGT splits into two terms: the CFIM and an
additional nonnegative measurement-transmitted metric in the phase sector.
Its imaginary part provides a semiclassical Berry curvature. Its loss
relative to the QGT is quantified by a measurement-dependent gap. Our
results provide an exactly solvable four-parameter platform for the
semiclassical geometric framework. They also clarify how realistic
measurements read out the geometric content of quantum states, and how they
partially degrade it.
\end{abstract}


\maketitle

\section{Introduction}

Fisher information offers a natural framework for distinguishing what is
operationally accessible through measurements from what is intrinsic to the
state itself \cite{1,2}. To quantify the information carried by a quantum
system, one typically studies a state that depends on a set of parameters
\cite{3,4}. The analysis relies on a measurement described by a positive
operator-valued measure (POVM) \cite{5}. The classical Fisher information
matrix (CFIM) quantifies how much information the outcome probabilities
carry about the parameters \cite{6,7}. The quantum Fisher information matrix
(QFIM), in contrast, sets the ultimate precision limit. This limit is
determined solely by the state \cite{8}. For every state and every POVM,
these two quantities (CFIM and QFIM) obey the cornerstone inequality
$\mathcal{F}_{C}\leq \mathcal{F}_{Q}$. In the single-parameter case, the
bound can always be saturated by an appropriate measurement. For the
multiple-parameter case, however, saturation is generically impossible
\cite{9,10,11}. The resulting gap between $\mathcal{F}_{C}$ and
$\mathcal{F}_{Q}$ is a genuinely quantum obstruction. It stems from the
noncommutativity of the measurements that are optimal for the different
parameters, not from a purely operational shortcoming \cite{12}.

From a geometric perspective, this gap is naturally encoded in the quantum
geometric tensor (QGT) \cite{13}. For a pure state, the QGT is a Hermitian,
positive semidefinite matrix. Its real part coincides with the QFIM. It also
defines the Fubini--Study metric on the state manifold. Its imaginary part
is the Berry curvature \cite{14}. A complete account of measurement
incompatibility in the multiparameter regime therefore demands a framework
beyond purely real metric structures \cite{15}. Indeed, the CFIM can
coincide with the QFIM only if the underlying curvature vanishes \cite{9,10}.
Moreover, the conditions for saturating the quantum Cram\'{e}r--Rao bound
have recently been organized into a hierarchy of commutativity conditions
\cite{16}.

To quantify the quantum obstruction in a measurement-dependent way, Imai et
al. recently introduced the semiclassical geometric tensor (SCGT) \cite{17}.
It is a gauge-invariant counterpart of the QGT. Unlike the QGT, it depends
on the chosen POVM. They proved the matrix inequality $\mathcal{C}\leq
\mathcal{Q}$, which sharpens $\mathcal{F}_{C}\leq \mathcal{F}_{Q}$. They
also showed that the real part of the SCGT decomposes into the CFIM and an
additional nonnegative contribution that captures the quantum obstruction.
Its imaginary part provides a semiclassical Berry curvature. Its difference
from the quantum curvature quantifies the curvature content lost by the
measurement. These ideas are closely connected with the mixed-state geometry
based on the Bures metric and the mean Uhlmann curvature \cite{18}. Related
recent work has also developed Fisher--Bures information geometry for
bosonic Gaussian thermal states \cite{19}.

In this work, we realize and extend the semiclassical geometric framework on
an exactly solvable four-parameter platform. The platform is a class of pure
four-parameter single-qutrit states (FPSQSs). These states are probed by a
one-parameter family of POVMs. The family interpolates between an
uninformative measurement and a sharp projective measurement. For this
family, we derive closed-form expressions for the CFIM, the QGT, and the
SCGT. We analyze their structural properties. We also establish the
projective and trivial limits of the SCGT. We show that the CFIM saturates
the QFIM in the angle sector at the projective limit. However, it remains
blind to the phase parameters for every measurement in the family. Moreover,
we quantify the loss of Berry curvature through the difference $\Delta
=\mathcal{G}-\mathcal{D}$. Our results provide a concrete multiparameter test
bed for the SCGT formalism. They also clarify the geometric content of
quantum states.

The paper is organized as follows. In Sec.~II we introduce the FPSQSs. In
Sec.~III we introduce the POVM family and compute the CFIM. In Sec.~IV we
derive the QGT and analyze its real and imaginary parts. In Sec.~V we
construct the SCGT and its decomposition. In Sec.~VI we quantify the
curvature gap $\Delta$. In Sec.~VII we compare the CFIM with the QFIM, and
we conclude in Sec.~VIII.

\section{Four-parameter single-qutrit states}

In this section, we introduce the class of pure four-parameter single-qutrit
states (FPSQSs) used throughout the paper, together with the notation
adopted below. A qutrit is a three-level quantum system, and qudit platforms
have attracted growing interest in quantum information processing
\cite{20,21}.

Here, we consider a class of pure qutrit states of the ket form
\begin{equation}
\left\vert \psi _{\mathbf{\lambda }}\right\rangle =s_{\theta }c_{\phi
}\left\vert 0\right\rangle +e^{i\mu }s_{\theta }s_{\phi }\left\vert
1\right\rangle +e^{i\nu }c_{\theta }\left\vert 2\right\rangle ,  \label{1.1}
\end{equation}
which encode a set of four real parameters $\mathbf{\lambda }=\left( \lambda
_{1},\lambda _{2},\lambda _{3},\lambda _{4}\right) =\left( \theta ,\phi ,\mu
,\nu \right) $. Among them, $\theta $ and $\phi $ are the angle parameters,
while $\mu $ and $\nu $ are the phase parameters. Moreover, we set their
intervals $\theta \in \lbrack 0,\pi ]$, $\phi \in \lbrack 0,2\pi ]$, $\mu
\in \lbrack 0,2\pi ]$, and $\nu \in \lbrack 0,2\pi ]$. For simplicity, we
set $s_{x}\equiv \sin x$ and $c_{x}\equiv \cos x$. We denote this class of
states by FPSQSs.

In the Hilbert space spanned by the basis $\{\left\vert 0\right\rangle
=\left(
\begin{array}{ccc}
1 & 0 & 0%
\end{array}%
\right) ^{T}$, $\left\vert 1\right\rangle =\left(
\begin{array}{ccc}
0 & 1 & 0%
\end{array}%
\right) ^{T}$, $\left\vert 2\right\rangle =\left(
\begin{array}{ccc}
0 & 0 & 1%
\end{array}%
\right) ^{T}$\}, the FPSQS takes the matrix form
\begin{equation}
\left\vert \psi _{\mathbf{\lambda }}\right\rangle =\left(
\begin{array}{ccc}
s_{\theta }c_{\phi } & e^{i\mu }s_{\theta }s_{\phi } & e^{i\nu }c_{\theta }%
\end{array}%
\right) ^{T},  \label{1.2}
\end{equation}
The bra form of $\left\vert \psi _{\mathbf{\lambda }}\right\rangle $ reads
\begin{equation}
\left\langle \psi _{\mathbf{\lambda }}\right\vert =\left(
\begin{array}{ccc}
s_{\theta }c_{\phi } & e^{-i\mu }s_{\theta }s_{\phi } & e^{-i\nu }c_{\theta }%
\end{array}%
\right) .  \label{1.3}
\end{equation}
The density operator $\rho _{\mathbf{\lambda }}=\left\vert \psi
_{\mathbf{\lambda }}\right\rangle \left\langle \psi _{\mathbf{\lambda }%
}\right\vert $ can be expressed as
\begin{equation}
\rho _{\mathbf{\lambda }}=\left(
\begin{array}{ccc}
s_{\theta }^{2}c_{\phi }^{2} & e^{-i\mu }s_{\theta }^{2}s_{2\phi }/2 &
e^{-i\nu }s_{2\theta }c_{\phi }/2 \\
e^{i\mu }s_{\theta }^{2}s_{2\phi }/2 & s_{\theta }^{2}s_{\phi }^{2} &
e^{i(\mu -\nu )}s_{2\theta }s_{\phi }/2 \\
e^{i\nu }s_{2\theta }c_{\phi }/2 & e^{-i(\mu -\nu )}s_{2\theta }s_{\phi }/2
& c_{\theta }^{2}%
\end{array}%
\right) .  \label{1.4}
\end{equation}

\section{Classical Fisher information matrix}

In this section, we study the classical Fisher information matrix (CFIM)
associated with a concrete three-outcome measurement. We introduce a family
of measurements, which can be adjusted by one parameter $\varepsilon $. The
CFIM quantifies how much information about the four parameters of the FPSQS
can be extracted from the measurement statistics.

\textit{POVM:} We take the following one-parameter POVM
\begin{equation}
\mathrm{E}\left( \varepsilon \right) =\{E_{\omega }\left( \varepsilon
\right) =\varepsilon \left\vert \omega \right\rangle \left\langle \omega
\right\vert +\left( 1-\varepsilon \right) \frac{\mathbb{I}}{3}\}  \label{2.1}
\end{equation}
with $\omega =0,1,2$, and a parameter $\varepsilon \in \left[ 0,1\right] $.
One can check that $E_{\omega }\left( \varepsilon \right) \succeq 0$ and
$\sum_{\omega =0}^{2}E_{\omega }\left( \varepsilon \right) =\mathbb{I}$.

The parameter $\varepsilon \in \left[ 0,1\right] $ tunes the selectivity of
the measurement. For $\varepsilon =1$, $\mathrm{E}\left( \varepsilon \right)
$ reduces to the sharp projective measurement onto the computational basis.
For $\varepsilon =0$, all three effects collapse to the same trivial
operator $\mathbb{I}/3$, whose outcome probabilities are state-independent
and therefore carry no information. For $0<\varepsilon <1$, $\mathrm{E}%
\left( \varepsilon \right) $ describes a partially selective measurement, in
which a fraction $\left( 1-\varepsilon \right) /3$ of every outcome is
replaced by an isotropic background. Since all effects are diagonal in the
computational basis $\{\left\vert 0\right\rangle ,\left\vert 1\right\rangle
,\left\vert 2\right\rangle \}$, the POVM is a measurement of the populations
of the three basis states. The probabilities $p_{\omega }\left( \mathbf{%
\lambda }\right) $ depend only on the squared amplitudes of $\left\vert \psi
_{\mathbf{\lambda }}\right\rangle $ and are insensitive to the phases $\mu $
and $\nu $. As we will see, this phase blindness is the main limitation of
the present measurement scheme.

Case (i): If $0<\varepsilon <1$, $\mathrm{E}\left( \varepsilon \right) $ is
a non-rank-one POVM with three outcomes
\begin{align}
E_{0}\left( \varepsilon \right) &=\left(
\begin{array}{ccc}
\frac{1+2\varepsilon }{3} & 0 & 0 \\
0 & \frac{1-\varepsilon }{3} & 0 \\
0 & 0 & \frac{1-\varepsilon }{3}%
\end{array}%
\right) ,  \notag \\
E_{1}\left( \varepsilon \right) &=\left(
\begin{array}{ccc}
\frac{1-\varepsilon }{3} & 0 & 0 \\
0 & \frac{1+2\varepsilon }{3} & 0 \\
0 & 0 & \frac{1-\varepsilon }{3}%
\end{array}%
\right) ,  \notag \\
E_{2}\left( \varepsilon \right) &=\left(
\begin{array}{ccc}
\frac{1-\varepsilon }{3} & 0 & 0 \\
0 & \frac{1-\varepsilon }{3} & 0 \\
0 & 0 & \frac{1+2\varepsilon }{3}%
\end{array}%
\right) .  \label{2.2}
\end{align}

Case (ii): If $\varepsilon =1$, $\mathrm{E}\left( \varepsilon \right) $
reduces to a rank-one POVM with three outcomes, $\mathrm{E}\left( 1\right)
=\{E_{\omega }\left( 1\right) =\left\vert \omega \right\rangle \left\langle
\omega \right\vert \}$ with $\omega =0,1,2$, i.e.,
\begin{align}
E_{0}\left( 1\right) &=\left(
\begin{array}{ccc}
1 & 0 & 0 \\
0 & 0 & 0 \\
0 & 0 & 0%
\end{array}%
\right) ,  \notag \\
E_{1}\left( 1\right) &=\left(
\begin{array}{ccc}
0 & 0 & 0 \\
0 & 1 & 0 \\
0 & 0 & 0%
\end{array}%
\right) ,  \notag \\
E_{2}\left( 1\right) &=\left(
\begin{array}{ccc}
0 & 0 & 0 \\
0 & 0 & 0 \\
0 & 0 & 1%
\end{array}%
\right) .  \label{2.3}
\end{align}
This case corresponds to a sharp projective measurement.

Case (iii): If $\varepsilon =0$, $\mathrm{E}\left( \varepsilon \right) $
reduces to a rank-three POVM in which all three outcomes coincide, $\mathrm{E%
}\left( 0\right) =\{E_{\omega }\left( 0\right) =\frac{\mathbb{I}}{3}\}$ with
$\omega =0,1,2$, i.e.,
\begin{align}
E_{0}\left( 0\right) &=E_{1}\left( 0\right) =E_{2}\left( 0\right)  \notag \\
&=\left(
\begin{array}{ccc}
\frac{1}{3} & 0 & 0 \\
0 & \frac{1}{3} & 0 \\
0 & 0 & \frac{1}{3}%
\end{array}%
\right) .  \label{2.4}
\end{align}
This case corresponds to an uninformative POVM.

\textit{Probabilities:} The outcome probabilities for the FPSQS under the
POVM above are obtained from $p_{\omega }=p_{\omega }\left( \mathbf{\lambda }
\right) =\left\langle \psi _{\mathbf{\lambda }}\right\vert E_{\omega
}\left\vert \psi _{\mathbf{\lambda }}\right\rangle $, so that we have
\begin{align}
p_{0} &=\frac{1-\varepsilon }{3}+\varepsilon s_{\theta }^{2}c_{\phi }^{2},
 \notag \\
p_{1} &=\frac{1-\varepsilon }{3}+\varepsilon s_{\theta }^{2}s_{\phi }^{2},
 \notag \\
p_{2} &=\frac{1-\varepsilon }{3}+\varepsilon c_{\theta }^{2}.  \label{2.5}
\end{align}
For $\varepsilon =1$ we have $p_{0}=s_{\theta }^{2}c_{\phi }^{2}$,
$p_{1}=s_{\theta }^{2}s_{\phi }^{2}$, and $p_{2}=c_{\theta }^{2}$; for
$\varepsilon =0$ we have $p_{0}=p_{1}=p_{2}=\frac{1}{3}$. In every case, one
readily checks that $p_{0}+p_{1}+p_{2}=1$.

The normalization $p_{0}+p_{1}+p_{2}=1$ follows directly from the
completeness of the POVM ($\sum_{\omega }E_{\omega }=\mathbb{I}$). It holds
for every $\varepsilon $. Each probability is a convex combination of two
ingredients. The first is the corresponding population ($s_{\theta
}^{2}c_{\phi }^{2}$, $s_{\theta }^{2}s_{\phi }^{2}$, $c_{\theta }^{2}$) of
the state. The second is the uniform background $1/3$. The weight
$\varepsilon $ measures how strongly the measurement outcome is correlated
with the state.

\textit{CFIM:} The CFIM characterizes the precision limit associated with
the chosen POVM. It is the Fisher information of the probability
distribution $\{p_{\omega }\left( \mathbf{\lambda }\right) \}$. Through the
classical Cram\'{e}r--Rao bound, it sets the ultimate precision achievable
by estimators constructed from the measurement outcomes \cite{6,7,22}. Since
the measurement itself is part of the estimation scheme, the CFIM depends on
the choice of $\mathrm{E}\left( \varepsilon \right) $. This contrasts with
the quantum Fisher information matrix (QFIM) of the state, which is
measurement independent. For any POVM, one has the general hierarchy
$\mathcal{F}_{C}\leq \mathcal{F}_{Q}$. We will make this hierarchy explicit
for the FPSQS below.

The CFIM $\mathcal{F}_{C}\left( \left\vert \psi _{\mathbf{\lambda }%
}\right\rangle \text{,}\mathrm{E}\left( \varepsilon \right) \right) $ is
defined as the matrix with elements
\begin{align}
\left( \mathcal{F}_{C}\right) _{jk} &=[\mathcal{F}_{C}\left( \left\vert
\psi _{\mathbf{\lambda }}\right\rangle \text{,}\mathrm{E}\left( \varepsilon
\right) \right) ]_{jk}  \notag \\
&=\sum_{\omega =0}^{2}\frac{[\partial _{j}p_{\omega }\left( \mathbf{\lambda
}\right) ][\partial _{k}p_{\omega }\left( \mathbf{\lambda }\right) ]}{%
p_{\omega }\left( \mathbf{\lambda }\right) }.  \label{2.6}
\end{align}
Taking the derivatives of the probabilities $p_{\omega }\left( \mathbf{%
\lambda }\right) $, we have
\begin{align*}
\partial _{1}p_{0} &=\varepsilon s_{2\theta }c_{\phi }^{2}, \qquad
\partial _{2}p_{0}=-\varepsilon s_{\theta }^{2}s_{2\phi }, \\
\partial _{3}p_{0} &=\partial _{4}p_{0}=0, \\
\partial _{1}p_{1} &=\varepsilon s_{2\theta }s_{\phi }^{2}, \qquad
\partial _{2}p_{1}=\varepsilon s_{\theta }^{2}s_{2\phi }, \\
\partial _{3}p_{1} &=\partial _{4}p_{1}=0, \\
\partial _{1}p_{2} &=-\varepsilon s_{2\theta }, \\
\partial _{2}p_{2} &=\partial _{3}p_{2}=\partial _{4}p_{2}=0.
\end{align*}
Thus, we can calculate the CFIM as follows.
\begin{equation}
\mathcal{F}_{C}\left( \left\vert \psi _{\mathbf{\lambda }}\right\rangle
\text{,}\mathrm{E}\left( \varepsilon \right) \right) =\left(
\begin{array}{cccc}
\left( \mathcal{F}_{C}\right) _{11} & \left( \mathcal{F}_{C}\right) _{12} & 0
& 0 \\
\left( \mathcal{F}_{C}\right) _{21} & \left( \mathcal{F}_{C}\right) _{22} & 0
& 0 \\
0 & 0 & 0 & 0 \\
0 & 0 & 0 & 0%
\end{array}%
\right)  \label{2.8}
\end{equation}
with
\begin{align*}
\left( \mathcal{F}_{C}\right) _{11} &=\varepsilon ^{2}s_{2\theta }^{2}\left(
\frac{c_{\phi }^{4}}{p_{0}}+\frac{s_{\phi }^{4}}{p_{1}}+\frac{1}{p_{2}}\right)
, \\
\left( \mathcal{F}_{C}\right) _{12} &=\left( \mathcal{F}_{C}\right)
_{21}=\varepsilon ^{2}s_{2\theta }s_{\theta }^{2}s_{2\phi }\left(\frac{s_{\phi
}^{2}}{p_{1}}-\frac{c_{\phi }^{2}}{p_{0}}\right), \\
\left( \mathcal{F}_{C}\right) _{22} &=\varepsilon ^{2}s_{\theta
}^{4}s_{2\phi }^{2}\left(\frac{1}{p_{0}}+\frac{1}{p_{1}}\right),
\end{align*}
and the remaining elements $\left( \mathcal{F}_{C}\right) _{jk}$ are zero.
Two structural features of this CFIM deserve emphasis. First, the CFIM
vanishes identically in the phase directions, $\left( \mathcal{F}_{C}\right)
_{33}=\left( \mathcal{F}_{C}\right) _{44}=\left( \mathcal{F}_{C}\right)
_{34}=0$. This is because the probabilities $p_{\omega }\left( \mathbf{\lambda
}\right) $ do not depend on $\mu $ and $\nu $. No population measurement can
reveal any information about the phases, which require measurements in a
coherent (off-diagonal) basis. Second, all matrix elements are proportional
to $\varepsilon ^{2}$. Since the signal $\partial _{j}p_{\omega }$ scales
with $\varepsilon $ while the background $\left( 1-\varepsilon \right) /3$
remains finite, the information carried by a weakly selective measurement is
suppressed quadratically. In the trivial limit $\varepsilon =0$, the CFIM
vanishes. For $0<\varepsilon \leq 1$, the CFIM has rank two (generically).
This reflects the fact that only the two amplitude parameters ($\theta $ and
$\phi $) are accessible to the measurement.

In particular, we have
\begin{equation}
\mathcal{F}_{C}\left( \left\vert \psi _{\mathbf{\lambda }}\right\rangle
\text{,}\mathrm{E}\left( 1\right) \right) =\left(
\begin{array}{cccc}
4 & 0 & 0 & 0 \\
0 & 4s_{\theta }^{2} & 0 & 0 \\
0 & 0 & 0 & 0 \\
0 & 0 & 0 & 0%
\end{array}%
\right) .  \label{2.10}
\end{equation}
Remarkably, this matrix coincides with the $\{\theta ,\phi \}$ sector of the
QFIM $\mathcal{F}_{Q}$, which will be derived in the next section. The
projective measurement onto the computational basis is optimal. In
particular, it saturates the quantum Cram\'{e}r--Rao bound for the
estimation of $\theta $ and $\phi $. Yet it remains completely blind to the
phases $\mu $ and $\nu $. This partial optimality, together with the phase
blindness, is exactly the content of the closeness analysis between the CFIM
and the QFIM. That analysis is carried out in the final part of the paper.
In the opposite limit, we have
\begin{equation}
\mathcal{F}_{C}\left( \left\vert \psi _{\mathbf{\lambda }}\right\rangle
\text{,}\mathrm{E}\left( 0\right) \right) =\left(
\begin{array}{cccc}
0 & 0 & 0 & 0 \\
0 & 0 & 0 & 0 \\
0 & 0 & 0 & 0 \\
0 & 0 & 0 & 0%
\end{array}%
\right) .  \label{2.11}
\end{equation}
As expected, a completely nonselective measurement yields no information
about any parameter. More generally, the $\varepsilon $ dependence of the
CFIM shows how the extractable information is controlled by the selectivity
of the measurement and by the background noise it introduces.

\section{Quantum geometric tensor}

The quantum geometric tensor (QGT) of a quantum state encodes the local
geometry of the state manifold in Hilbert space \cite{23,24,25}. Its real
part defines the quantum metric (or Fubini--Study metric), while its
imaginary part defines the Berry curvature \cite{15}. In this section we
compute the QGT of $\left\vert \psi _{\mathbf{\lambda }}\right\rangle $
following the approach of Tan et al. \cite{26}.

The QGT $\mathcal{Q}\left( \left\vert \psi _{\mathbf{\lambda }}\right\rangle
\right) $ of $\left\vert \psi _{\mathbf{\lambda }}\right\rangle $ is the
matrix with elements
\begin{equation}
\lbrack \mathcal{Q}\left( \left\vert \psi _{\mathbf{\lambda }}\right\rangle
\right) ]_{jk}=\mathrm{tr}[\rho _{\mathbf{\lambda }}L_{j}\left( \mathbf{%
\lambda }\right) L_{k}\left( \mathbf{\lambda }\right) ],  \label{3.1}
\end{equation}
where $L_{j}\left( \mathbf{\lambda }\right) $ is the symmetric logarithmic
derivative (SLD) of $\left\vert \psi _{\mathbf{\lambda }}\right\rangle $
defined as
\begin{equation}
L_{j}\left( \mathbf{\lambda }\right) =2\left( \left\vert \partial _{j}\psi
_{\mathbf{\lambda }}\right\rangle \left\langle \psi _{\mathbf{\lambda }%
}\right\vert +\left\vert \psi _{\mathbf{\lambda }}\right\rangle \left\langle
\partial _{j}\psi _{\mathbf{\lambda }}\right\vert \right) .  \label{3.2}
\end{equation}
with $\left\vert \partial _{j}\psi _{\mathbf{\lambda }}\right\rangle
=\partial \left\vert \psi _{\mathbf{\lambda }}\right\rangle /\partial
\lambda _{j}$. In an alternative and often more convenient form,
$[\mathcal{Q}\left( \left\vert \psi _{\mathbf{\lambda }}\right\rangle
\right) ]_{jk}$ can also be calculated by projecting the derivatives onto
the subspace orthogonal to $\left\vert \psi _{\mathbf{\lambda }}\right\rangle
$
\begin{equation}
\lbrack \mathcal{Q}\left( \left\vert \psi _{\mathbf{\lambda }}\right\rangle
\right) ]_{jk}=4\left\langle \partial _{j}\psi _{\mathbf{\lambda }%
}\right\vert \left( \mathbb{I}-\left\vert \psi _{\mathbf{\lambda }%
}\right\rangle \left\langle \psi _{\mathbf{\lambda }}\right\vert \right)
\left\vert \partial _{k}\psi _{\mathbf{\lambda }}\right\rangle .  \label{3.3}
\end{equation}

For $\left\vert \psi _{\mathbf{\lambda }}\right\rangle $, we have
\begin{align*}
\left\vert \partial _{1}\psi _{\mathbf{\lambda }}\right\rangle &=\left(
\begin{array}{ccc}
c_{\theta }c_{\phi } & e^{i\mu }c_{\theta }s_{\phi } & -e^{i\nu }s_{\theta }%
\end{array}%
\right) ^{T}, \\
\left\vert \partial _{2}\psi _{\mathbf{\lambda }}\right\rangle &=\left(
\begin{array}{ccc}
-s_{\theta }s_{\phi } & e^{i\mu }s_{\theta }c_{\phi } & 0%
\end{array}%
\right) ^{T}, \\
\left\vert \partial _{3}\psi _{\mathbf{\lambda }}\right\rangle &=\left(
\begin{array}{ccc}
0 & ie^{i\mu }s_{\theta }s_{\phi } & 0%
\end{array}%
\right) ^{T}, \\
\left\vert \partial _{4}\psi _{\mathbf{\lambda }}\right\rangle &=\left(
\begin{array}{ccc}
0 & 0 & ie^{i\nu }c_{\theta }%
\end{array}%
\right) ^{T},
\end{align*}
together with
\begin{align*}
\left\langle \partial _{1}\psi _{\mathbf{\lambda }}\right\vert &=\left(
\begin{array}{ccc}
c_{\theta }c_{\phi } & e^{-i\mu }c_{\theta }s_{\phi } & -e^{-i\nu }s_{\theta
}%
\end{array}%
\right) , \\
\left\langle \partial _{2}\psi _{\mathbf{\lambda }}\right\vert &=\left(
\begin{array}{ccc}
-s_{\theta }s_{\phi } & e^{-i\mu }s_{\theta }c_{\phi } & 0%
\end{array}%
\right) , \\
\left\langle \partial _{3}\psi _{\mathbf{\lambda }}\right\vert &=\left(
\begin{array}{ccc}
0 & -ie^{-i\mu }s_{\theta }s_{\phi } & 0%
\end{array}%
\right) , \\
\left\langle \partial _{4}\psi _{\mathbf{\lambda }}\right\vert &=\left(
\begin{array}{ccc}
0 & 0 & -ie^{-i\nu }c_{\theta }%
\end{array}%
\right) .
\end{align*}
A direct calculation gives
\begin{equation*}
L_{1}\left( \mathbf{\lambda }\right) =\left(
\begin{array}{ccc}
2s_{2\theta }c_{\phi }^{2} & e^{-i\mu }s_{2\theta }s_{2\phi } & 2e^{-i\nu
}c_{2\theta }c_{\phi } \\
e^{i\mu }s_{2\theta }s_{2\phi } & 2s_{2\theta }s_{\phi }^{2} & 2e^{i(\mu
-\nu )}c_{2\theta }s_{\phi } \\
2e^{i\nu }c_{2\theta }c_{\phi } & 2e^{-i(\mu -\nu )}c_{2\theta }s_{\phi } &
-2s_{2\theta }%
\end{array}%
\right) ,
\end{equation*}
\begin{equation*}
L_{2}\left( \mathbf{\lambda }\right) =\left(
\begin{array}{ccc}
-2s_{\theta }^{2}s_{2\phi } & 2e^{-i\mu }s_{\theta }^{2}c_{2\phi } &
-e^{-i\nu }s_{2\theta }s_{\phi } \\
2e^{i\mu }s_{\theta }^{2}c_{2\phi } & 2s_{\theta }^{2}s_{2\phi } & e^{i(\mu
-\nu )}s_{2\theta }c_{\phi } \\
-e^{i\nu }s_{2\theta }s_{\phi } & e^{-i(\mu -\nu )}s_{2\theta }c_{\phi } & 0%
\end{array}%
\right) ,
\end{equation*}
\begin{equation*}
L_{3}\left( \mathbf{\lambda }\right) =\left(
\begin{array}{ccc}
0 & -ie^{-i\mu }s_{\theta }^{2}s_{2\phi } & 0 \\
ie^{i\mu }s_{\theta }^{2}s_{2\phi } & 0 & ie^{i(\mu -\nu )}s_{2\theta
}s_{\phi } \\
0 & -ie^{-i(\mu -\nu )}s_{2\theta }s_{\phi } & 0%
\end{array}%
\right) ,
\end{equation*}
\begin{equation*}
L_{4}\left( \mathbf{\lambda }\right) =\left(
\begin{array}{ccc}
0 & 0 & -ie^{-i\nu }s_{2\theta }c_{\phi } \\
0 & 0 & -ie^{i(\mu -\nu )}s_{2\theta }s_{\phi } \\
ie^{i\nu }s_{2\theta }c_{\phi } & ie^{-i(\mu -\nu )}s_{2\theta }s_{\phi } & 0%
\end{array}%
\right) .
\end{equation*}

Using Eq.~\eqref{3.1} or Eq.~\eqref{3.3}, we have
\begin{equation}
\arraycolsep=3pt
\mathcal{Q}\left( \left\vert \psi _{\mathbf{\lambda }}\right\rangle \right)
={\small \left(
\begin{array}{cccc}
4 & 0 & 2is_{2\theta }s_{\phi }^{2} & -2is_{2\theta } \\
0 & 4s_{\theta }^{2} & 2is_{\theta }^{2}s_{2\phi } & 0 \\
-2is_{2\theta }s_{\phi }^{2} & -2is_{\theta }^{2}s_{2\phi } & 4s_{\theta
}^{2}s_{\phi }^{2}-4s_{\theta }^{4}s_{\phi }^{4} & -s_{2\theta }^{2}s_{\phi
}^{2} \\
2is_{2\theta } & 0 & -s_{2\theta }^{2}s_{\phi }^{2} & s_{2\theta }^{2}%
\end{array}%
\right) .}\label{3.8}
\end{equation}
A few structural observations about the QGT in Eq.~\eqref{3.8} are in
order. As a Hermitian matrix, $\mathcal{Q}$ admits a unique decomposition
into its real and imaginary parts, $\mathcal{Q}=\mathcal{F}_{Q}+i\mathcal{G}$.
Here, the real symmetric part $\mathcal{F}_{Q}$ is the quantum metric. For
pure states, this is equivalent to the QFIM. The imaginary part,
represented by the real antisymmetric matrix $\mathcal{G}$, is the Berry
curvature. These two parts will be examined in turn below. For the FPSQS,
the QGT is remarkably sparse. The $\theta $--$\phi $ entry $\mathcal{Q}_{12}$
and the $\phi $--$\nu $ entry $\mathcal{Q}_{24}$ vanish identically, whereas
the $\mu $--$\nu $ entry $\mathcal{Q}_{34}$ is real and generically nonzero;
only its imaginary part vanishes. Every element is independent of the phases
$\mu $ and $\nu $. The nonvanishing imaginary elements ($\mathcal{Q}%
_{13}=2is_{2\theta }s_{\phi }^{2}$, $\mathcal{Q}_{14}=-2is_{2\theta }$, and
$\mathcal{Q}_{23}=2is_{\theta }^{2}s_{2\phi }$) couple each phase direction
to one of the angles. They encode the geometric phase content of the family,
which, as we shall see, lives entirely in the mixed angle-phase sectors. The
diagonal entries $\mathcal{Q}_{33}=4s_{\theta }^{2}s_{\phi }^{2}(1-s_{\theta
}^{2}s_{\phi }^{2})$ and $\mathcal{Q}_{44}=s_{2\theta }^{2}$ measure how
strongly the phases $\mu $ and $\nu $ displace the state. Both
$\mathcal{Q}_{33}$ and $\mathcal{Q}_{44}$ vanish at the degenerate points, at
which the corresponding amplitude disappears and the parametrization becomes
redundant. Finally, because the QGT is a positive semidefinite Hermitian
matrix, its real and imaginary parts are not independent. For every pair of
parameters, they obey the geometric uncertainty relation $[\mathcal{G}]%
_{jk}^{2}\leq \lbrack \mathcal{F}_{Q}]_{jj}[\mathcal{F}_{Q}]_{kk}-[\mathcal{F}%
_{Q}]_{jk}^{2}$. This relation bounds the Berry curvature in any
two-parameter plane by the corresponding metric block. It also ties the two
objects together at the degenerate points.

\subsection{Real part of the QGT}

The real part of the QGT is the quantum metric, i.e., the Fubini--Study
metric of the state manifold \cite{13,27}. For pure states, this real part
coincides exactly with the QFIM, i.e.,
\begin{align}
\mathcal{F}_{Q}\left( \left\vert \psi _{\mathbf{\lambda }}\right\rangle
\right) &=\text{Re}[\mathcal{Q}\left( \left\vert \psi _{\mathbf{\lambda }%
}\right\rangle \right) ]  \notag \\
&=\left(
\begin{array}{cccc}
4 & 0 & 0 & 0 \\
0 & 4s_{\theta }^{2} & 0 & 0 \\
0 & 0 & 4s_{\theta }^{2}s_{\phi }^{2}-4s_{\theta }^{4}s_{\phi }^{4} &
-s_{2\theta }^{2}s_{\phi }^{2} \\
0 & 0 & -s_{2\theta }^{2}s_{\phi }^{2} & s_{2\theta }^{2}%
\end{array}%
\right) .  \label{3.9}
\end{align}
The quantum metric defines an intrinsic notion of distance between
neighboring quantum states in the parameter space. The distance between
$\left\vert \psi _{\mathbf{\lambda }}\right\rangle $ and $\left\vert \psi
_{\mathbf{\lambda +d\lambda }}\right\rangle $ is obtained from
\begin{align}
ds^{2} &=1-\left\vert \left\langle \psi _{\mathbf{\lambda }}\right\vert
\left\vert \psi _{\mathbf{\lambda +d\lambda }}\right\rangle \right\vert ^{2}
 \notag \\
&=\frac{1}{4}\sum_{jk}[\mathcal{F}_{Q}\left( \left\vert \psi _{\mathbf{%
\lambda }}\right\rangle \right) ]_{jk}d\lambda _{j}d\lambda _{k}
\label{3.10}
\end{align}
This distance is the infinitesimal version of the statistical distance
between probability distributions \cite{28,29}. It controls the
leading-order decay of the fidelity between neighboring states,
$\left\vert \left\langle \psi _{\mathbf{\lambda }}\right\vert \left\vert
\psi _{\mathbf{\lambda +d\lambda }}\right\rangle \right\vert ^{2}\simeq
1-ds^{2}$. Hence, it quantifies how many copies of the state are required to
statistically resolve two nearby parameter values.

Physically, the real part plays two further roles. (i) \textit{Quantum
metrology}. As the QFIM, it determines the ultimate precision with which the
four parameters can be estimated. For any POVM, the resulting CFIM is
bounded from above by the QFIM, $\mathcal{F}_{C}\leq \mathcal{F}_{Q}$.
Equality is attained only by informationally optimal measurements. Recent
work has further connected QFI-matrix elements to entanglement monotones and
achievable precision in multiparameter estimation \cite{30}. (ii)
\textit{Quantum speed limit}. The metric also bounds the speed of quantum
evolution. According to the Anandan--Aharonov relation \cite{31}, the
Fubini--Study distance travelled per unit time cannot exceed twice the
energy uncertainty of the state. Hence, the quantum metric controls the
minimal time required for the state to evolve into a distinguishable
configuration.

For the FPSQS, the QFIM exhibits a clear structure. First, all its matrix
elements depend only on the angles $\theta $ and $\phi $, not on the phase
parameters $\mu $ and $\nu $. The $\{\mu ,\nu \}$ sector therefore describes
a flat torus whose shape is modulated by $\{\theta ,\phi \}$. Second, the
QFIM is block diagonal. The angle sector is thus metrically decoupled from
the phase sector. In particular, there is no $\theta $--$\phi $ cross term
in the metric. Third, the determinant of the $\{\mu ,\nu \}$ block reads
$16s_{\theta }^{4}c_{\theta }^{2}s_{\phi }^{2}c_{\phi }^{2}\geq 0$. It
vanishes whenever $\theta =0$, $\theta =\pi /2$, $\phi =0$, or $\phi =\pi /2$
(mod $\pi $). At these points, the state loses its dependence on some of the
parameters, and the parametrization becomes redundant. Consequently, the
QFIM drops rank, and neighboring states become indistinguishable along the
degenerate directions. Such degeneracies signal the singular behavior of the
state manifold. They are widely used to detect quantum phase transitions, at
which the quantum metric diverges or loses rank \cite{15,27,32}.

\subsection{Imaginary part of the QGT}

The imaginary part of the QGT is the Berry curvature of the state manifold
\cite{14}:
\begin{align}
\mathcal{G}\left( \left\vert \psi _{\mathbf{\lambda }}\right\rangle \right)
&=\text{Im}[\mathcal{Q}\left( \left\vert \psi _{\mathbf{\lambda }%
}\right\rangle \right) ]  \notag \\
&=\left(
\begin{array}{cccc}
0 & 0 & 2s_{2\theta }s_{\phi }^{2} & -2s_{2\theta } \\
0 & 0 & 2s_{\theta }^{2}s_{2\phi } & 0 \\
-2s_{2\theta }s_{\phi }^{2} & -2s_{\theta }^{2}s_{2\phi } & 0 & 0 \\
2s_{2\theta } & 0 & 0 & 0%
\end{array}%
\right)  \label{3.11}
\end{align}
It governs the geometric (Berry) phase accumulated by the state under an
adiabatic cyclic evolution in the parameter space. One first introduces the
Berry connection $A_{j}\left( \mathbf{\lambda }\right) =i\left\langle \psi
_{\mathbf{\lambda }}\right\vert \left. \partial _{j}\psi _{\mathbf{\lambda }%
}\right\rangle $. Then, by Stokes' theorem, the Berry phase along a closed
loop $C$ bounding a surface $S$ is given as
\begin{align}
\gamma _{B} &=\oint_{C}A_{j}\left( \mathbf{\lambda }\right) d\lambda _{j}
 \notag \\
&=-\frac{1}{2}\int_{S}\sum_{j<k}[\mathcal{G}\left( \left\vert \psi
_{\mathbf{\lambda }}\right\rangle \right) ]_{jk}\,d\lambda _{j}\wedge d\lambda
_{k}.  \label{3.12}
\end{align}
Here, the prefactor $-1/2$ follows from the normalization of $\mathcal{G}$
in Eq.~\eqref{3.11}. Unlike the connection, the curvature $\mathcal{G}$ is
gauge invariant. In particular, it is independent of the phase convention of
$\left\vert \psi _{\mathbf{\lambda }}\right\rangle $. It depends only on the
manifold traced out by the state. Moreover, $\mathcal{G}$ vanishes
identically for a single parameter. This is why at least two parameters are
required to observe a nontrivial geometric phase.

The Berry curvature has two particularly important physical consequences.
(i) \textit{Topology}. The integral of $\mathcal{G}$ over any closed
two-dimensional submanifold of the parameter space is a quantized
topological invariant, namely a Chern number. It characterizes the global
twisting of the state bundle. Such invariants underlie the geometric
response of physical systems, for example the anomalous Hall effect and
topological photonic transport \cite{15,33,34}. (ii) \textit{Multiparameter
quantum metrology}. The imaginary part of the QGT is the Uhlmann curvature,
which appears in the multiparameter estimation problem. A nonvanishing
curvature makes the symmetric logarithmic derivatives noncommuting on
average. Hence, the parameters cannot be estimated simultaneously at the
ultimate quantum precision limit. The achievable precision is then governed
by the Holevo bound rather than by the QFIM alone \cite{4,35}.

For the FPSQS, the Berry curvature also depends only on $\theta $ and $\phi
$. Its only nonvanishing independent components are $[\mathcal{G}%
]_{13}=2s_{2\theta }s_{\phi }^{2}$, $[\mathcal{G}]_{14}=-2s_{2\theta }$, and
$[\mathcal{G}]_{23}=2s_{\theta }^{2}s_{2\phi }$ (together with their
antisymmetric partners), whereas $[\mathcal{G}]_{12}=[\mathcal{G}]_{24}=[%
\mathcal{G}]_{34}=0$. Hence the $\{\theta ,\phi \}$ and $\{\mu ,\nu \}$
planes are each individually curvature-free. Since $\mathcal{G}$ does not
vanish generically, the four parameters of the FPSQS are intrinsically
incompatible for simultaneous optimal estimation. This incompatibility
motivates the semiclassical analysis in the following sections.

\section{Semiclassical geometric tensor}

In analogy with the QGT in Eq.~\eqref{3.8}, the semiclassical geometric
tensor (SCGT) \cite{17} of $\left\vert \psi _{\mathbf{\lambda }%
}\right\rangle $ associated with $\mathrm{E}\left( \varepsilon \right) $ is
the matrix $\mathcal{C}\left( \left\vert \psi _{\mathbf{\lambda }%
}\right\rangle ,\mathrm{E}\left( \varepsilon \right) \right) $ with elements
\begin{align}
&[\mathcal{C}\left( \left\vert \psi _{\mathbf{\lambda }}\right\rangle
\text{,}\mathrm{E}\left( \varepsilon \right) \right) ]_{jk}  \notag \\
&\quad=4\left\langle \partial _{j}\psi _{\mathbf{\lambda }}\right\vert \left(
\mathcal{M}\left( \left\vert \psi _{\mathbf{\lambda }}\right\rangle
\text{,}\mathrm{E}\left( \varepsilon \right) \right) -\left\vert \psi
_{\mathbf{\lambda }}\right\rangle \left\langle \psi _{\mathbf{\lambda }%
}\right\vert \right) \left\vert \partial _{k}\psi _{\mathbf{\lambda }%
}\right\rangle
\label{4.1}
\end{align}
where
\begin{equation}
\mathcal{M}\left( \left\vert \psi _{\mathbf{\lambda }}\right\rangle
\text{,}\mathrm{E}\left( \varepsilon \right) \right) =\sum_{\omega
=0}^{2}\frac{E_{\omega }\left\vert \psi _{\mathbf{\lambda }}\right\rangle
\left\langle \psi _{\mathbf{\lambda }}\right\vert E_{\omega }}{p_{\omega
}\left( \mathbf{\lambda }\right) }.  \label{4.2}
\end{equation}
Equivalently, $[\mathcal{C}\left( \left\vert \psi _{\mathbf{\lambda }%
}\right\rangle \text{,}\mathrm{E}\left( \varepsilon \right) \right) ]_{jk}$
can be expressed in terms of the individual outcome responses as
\begin{align}
\lbrack \mathcal{C}\left( \left\vert \psi _{\mathbf{\lambda }}\right\rangle
\text{,}\mathrm{E}\left( \varepsilon \right) \right) ]_{jk} &=\mathcal{C}%
_{jk}  \notag \\
&=\sum_{\omega =0}^{2}\frac{\chi _{\omega ,j}^{\ast }\left( \mathbf{%
\lambda }\right) \chi _{\omega ,k}\left( \mathbf{\lambda }\right) }{%
p_{\omega }\left( \mathbf{\lambda }\right) }.  \label{4.3}
\end{align}
where $\chi _{\omega ,j}\left( \mathbf{\lambda }\right) =\left\langle \psi
_{\mathbf{\lambda }}\right\vert E_{\omega }L_{j}\left( \mathbf{\lambda }%
\right) \left\vert \psi _{\mathbf{\lambda }}\right\rangle $ and $\chi
_{\omega ,j}^{\ast }\left( \mathbf{\lambda }\right) $ is the complex
conjugate of $\chi _{\omega ,j}\left( \mathbf{\lambda }\right) $. Each
$\chi _{\omega ,j}$ measures how the $\omega $-th outcome responds to an
infinitesimal change of $\lambda _{j}$. Indeed, $\chi _{\omega ,j}$ is the
expectation value of the measurement-modified SLD $E_{\omega }L_{j}$ in the
state $\left\vert \psi _{\mathbf{\lambda }}\right\rangle $. Written as
$\mathcal{C}=\sum_{\omega }p_{\omega }^{-1}\left\vert \chi _{\omega
}\right\rangle \left\langle \chi _{\omega }\right\vert $ with $\left( \chi
_{\omega }\right) _{j}=\chi _{\omega ,j}$, the SCGT is a Gram-type matrix,
i.e., a sum of rank-one positive semidefinite matrices weighted by the
inverse outcome probabilities. Therefore, the SCGT is Hermitian and positive
semidefinite, with a symmetric real part and an antisymmetric imaginary
part. Moreover, the phases ($\mu $ and $\nu $) cancel in the products $\chi
_{\omega ,j}^{\ast }\chi _{\omega ,k}$. Hence, exactly as for the QGT, the
SCGT $\mathcal{C}$ depends only on the angles ($\theta $ and $\phi $).

For the FPSQS and the POVM effects above, we calculate $\chi _{\omega
,i}\left( \mathbf{\lambda }\right) $ as
\begin{align*}
\chi _{0,1} &=\varepsilon s_{2\theta }c_{\phi }^{2}, \qquad
\chi _{0,2}=-\varepsilon s_{\theta }^{2}s_{2\phi }, \\
\chi _{0,3} &=-i\frac{\varepsilon }{2}s_{\theta }^{4}s_{2\phi }^{2}, \qquad
\chi _{0,4}=-i\frac{\varepsilon }{2}s_{2\theta }^{2}c_{\phi }^{2}, \\
\chi _{1,1} &=\varepsilon s_{2\theta }s_{\phi }^{2}, \qquad
\chi _{1,2}=\varepsilon s_{\theta }^{2}s_{2\phi }, \\
\chi _{1,3} &=i\frac{\varepsilon }{2}\left(s_{2\theta }^{2}s_{\phi
}^{2}+s_{\theta }^{4}s_{2\phi }^{2}\right), \qquad
\chi _{1,4}=-i\frac{\varepsilon }{2}s_{2\theta }^{2}s_{\phi }^{2}, \\
\chi _{2,1} &=-\varepsilon s_{2\theta }, \qquad
\chi _{2,2}=0, \\
\chi _{2,3} &=-i\frac{\varepsilon }{2}s_{2\theta }^{2}s_{\phi }^{2}, \qquad
\chi _{2,4}=i\frac{\varepsilon }{2}s_{2\theta }^{2}.
\end{align*}

Thus, we have
\begin{equation}
\mathcal{C}\left( \left\vert \psi _{\mathbf{\lambda }}\right\rangle
\text{,}\mathrm{E}\left( \varepsilon \right) \right) =\left(
\begin{array}{cccc}
\mathcal{C}_{11} & \mathcal{C}_{12} & \mathcal{C}_{13} & \mathcal{C}_{14} \\
\mathcal{C}_{21} & \mathcal{C}_{22} & \mathcal{C}_{23} & \mathcal{C}_{24} \\
\mathcal{C}_{31} & \mathcal{C}_{32} & \mathcal{C}_{33} & \mathcal{C}_{34} \\
\mathcal{C}_{41} & \mathcal{C}_{42} & \mathcal{C}_{43} & \mathcal{C}_{44}%
\end{array}%
\right)  \label{4.4}
\end{equation}
where
\begin{align*}
\mathcal{C}_{11} &=\varepsilon ^{2}s_{2\theta }^{2}\left(\frac{c_{\phi }^{4}}{%
p_{0}}+\frac{s_{\phi }^{4}}{p_{1}}+\frac{1}{p_{2}}\right), \\
\mathcal{C}_{22} &=\varepsilon ^{2}s_{\theta }^{4}s_{2\phi }^{2}\left(\frac{1}{%
p_{0}}+\frac{1}{p_{1}}\right), \\
\mathcal{C}_{33} &=\frac{\varepsilon ^{2}s_{\theta }^{8}s_{2\phi }^{4}}{%
4p_{0}}+\frac{\varepsilon ^{2}\left(s_{2\theta }^{2}s_{\phi }^{2}+s_{\theta
}^{4}s_{2\phi }^{2}\right)^{2}}{4p_{1}}+\frac{\varepsilon ^{2}s_{2\theta
}^{4}s_{\phi }^{4}}{4p_{2}}, \\
\mathcal{C}_{44} &=\frac{\varepsilon ^{2}s_{2\theta }^{4}}{4}\left(\frac{c_{\phi
}^{4}}{p_{0}}+\frac{s_{\phi }^{4}}{p_{1}}+\frac{1}{p_{2}}\right),
\end{align*}
and
\begin{align*}
\mathcal{C}_{12} &=\mathcal{C}_{21}=\varepsilon ^{2}s_{2\theta }s_{\theta
}^{2}s_{2\phi }\left(\frac{s_{\phi }^{2}}{p_{1}}-\frac{c_{\phi }^{2}}{p_{0}}\right),
\\
\mathcal{C}_{34} &=\mathcal{C}_{43} \\
&=4\varepsilon ^{2}c_{\theta }^{2}s_{\theta }^{4}s_{\phi }^{2}\left(\frac{%
s_{\theta }^{2}c_{\phi }^{4}}{p_{0}}-\frac{s_{\phi
}^{2}\left(c_{\theta }^{2}+c_{\phi }^{2}s_{\theta }^{2}\right)}{p_{1}}-\frac{c_{\theta
}^{2}}{p_{2}}\right)
\end{align*}
as well as
\begin{align*}
\mathcal{C}_{13} &=\mathcal{C}_{31}^{\ast }=2i\varepsilon ^{2}s_{2\theta
}s_{\theta }^{2}s_{\phi }^{2}\left(\frac{c_{\theta }^{2}s_{\phi }^{2}+s_{\theta
}^{2}s_{\phi }^{2}c_{\phi }^{2}}{p_{1}}+\frac{c_{\theta }^{2}}{%
p_{2}}-\frac{s_{\theta }^{2}c_{\phi }^{4}}{p_{0}}\right), \\
\mathcal{C}_{14} &=\mathcal{C}_{41}^{\ast }=-i\frac{\varepsilon
^{2}s_{2\theta }^{3}}{2}\left(\frac{c_{\phi }^{4}}{p_{0}}+\frac{s_{\phi }^{4}}{%
p_{1}}+\frac{1}{p_{2}}\right), \\
\mathcal{C}_{23} &=\mathcal{C}_{32}^{\ast }=2i\varepsilon ^{2}s_{2\phi
}s_{\phi }^{2}s_{\theta }^{4}\left(\frac{s_{\theta }^{2}c_{\phi }^{2}}{p_{0}}+%
\frac{c_{\theta }^{2}+s_{\theta }^{2}c_{\phi }^{2}}{p_{1}}\right), \\
\mathcal{C}_{24} &=\mathcal{C}_{42}^{\ast }=i\frac{\varepsilon ^{2}}{2}%
s_{\theta }^{2}s_{2\theta }^{2}s_{2\phi }\left(\frac{c_{\phi }^{2}}{p_{0}}-\frac{%
s_{\phi }^{2}}{p_{1}}\right).
\end{align*}

Note that the elements $\mathcal{C}_{11}$, $\mathcal{C}_{22}$, $\mathcal{C}%
_{33}$, $\mathcal{C}_{44}$, $\mathcal{C}_{12}=\mathcal{C}_{21}$, and
$\mathcal{C}_{34}=\mathcal{C}_{43}$ are real, whereas $\mathcal{C}_{13}=%
\mathcal{C}_{31}^{\ast }$, $\mathcal{C}_{14}=\mathcal{C}_{41}^{\ast }$, $
\mathcal{C}_{23}=\mathcal{C}_{32}^{\ast }$, and $\mathcal{C}_{24}=\mathcal{C}%
_{42}^{\ast }$ are purely imaginary. Obviously, $\mathcal{C}$ is Hermitian
because of $\mathcal{C}_{jk}=\mathcal{C}_{kj}^{\ast }$. The real part of
$\mathcal{C}$ therefore collects only the $\{\theta ,\phi \}$ and $\{\mu ,\nu
\}$ blocks, while the imaginary part lives entirely in the mixed angle-phase
sectors, exactly as for the QGT. Every element of $\mathcal{C}$ is
proportional to $\varepsilon ^{2}$ and independent of $\mu $ and $\nu $. The
two extreme limits are instructive. For $\varepsilon =1$, the POVM reduces
to the sharp projective measurement $E_{\omega }=\left\vert \omega
\right\rangle \left\langle \omega \right\vert $, for which the operator
$\mathcal{M}$ introduced above reduces to $\sum_{\omega }\left\vert \omega
\right\rangle \left\langle \omega \right\vert =\mathbb{I}$. The SCGT then
coincides exactly with the QGT of the pure state, $\mathcal{C}\left(
\varepsilon =1\right) =\mathcal{Q}$, as is clear from Eqs.~\eqref{3.8} and
\eqref{4.4}. For $\varepsilon =0$, the POVM is the trivial uninformative
measurement $E_{\omega }=\mathbb{I}/3$, for which $\mathcal{M}=\left\vert
\psi _{\mathbf{\lambda }}\right\rangle \left\langle \psi _{\mathbf{\lambda }%
}\right\vert $ and the SCGT vanishes identically, $\mathcal{C}\left(
\varepsilon =0\right) =0$. In between, the SCGT scales as $\varepsilon ^{2}$
and interpolates smoothly between these two extremes.

\subsection{Real part of the SCGT}

The real part of $\mathcal{C}\left( \left\vert \psi _{\mathbf{\lambda }%
}\right\rangle \text{,}\mathrm{E}\left( \varepsilon \right) \right) $ can be
written as
\begin{align}
\text{Re}[\mathcal{C}\left( \left\vert \psi _{\mathbf{\lambda }%
}\right\rangle \text{,}\mathrm{E}\left( \varepsilon \right) \right) ]&=%
\mathcal{F}_{C}\left( \left\vert \psi _{\mathbf{\lambda }}\right\rangle
\text{,}\mathrm{E}\left( \varepsilon \right) \right) +\mathcal{I}\left(
\left\vert \psi _{\mathbf{\lambda }}\right\rangle \text{,}\mathrm{E}\left(
\varepsilon \right) \right) ,  \label{4.5}
\end{align}
Because the mixed angle-phase elements $\mathcal{C}_{13}$, $\mathcal{C}_{14}$%
, $\mathcal{C}_{23}$, and $\mathcal{C}_{24}$ are purely imaginary, the real
part keeps only the two diagonal blocks and takes the block-diagonal form
\begin{align}
\text{Re}[\mathcal{C}\left( \left\vert \psi _{\mathbf{\lambda }%
}\right\rangle \text{,}\mathrm{E}\left( \varepsilon \right) \right) ]&=\left(
\begin{array}{cccc}
\mathcal{C}_{11} & \mathcal{C}_{12} & 0 & 0 \\
\mathcal{C}_{21} & \mathcal{C}_{22} & 0 & 0 \\
0 & 0 & \mathcal{C}_{33} & \mathcal{C}_{34} \\
0 & 0 & \mathcal{C}_{43} & \mathcal{C}_{44}%
\end{array}%
\right) .  \label{4.6}
\end{align}

\subsubsection{The classical part $\mathcal{F}_{C}$}

Interestingly, we find that $\mathcal{C}_{11}=\left( \mathcal{F}_{C}\right)
_{11}$, $\mathcal{C}_{12}=\left( \mathcal{F}_{C}\right) _{12}$, $\mathcal{C}%
_{21}=\left( \mathcal{F}_{C}\right) _{21}$, and $\mathcal{C}_{22}=\left(
\mathcal{F}_{C}\right) _{22}$. This part is just the CFIM in Eq.~\eqref{2.8}.
That is, we have
\begin{equation}
\mathcal{F}_{C}\left( \left\vert \psi _{\mathbf{\lambda }}\right\rangle
\text{,}\mathrm{E}\left( \varepsilon \right) \right) =\left(
\begin{array}{cccc}
\mathcal{C}_{11} & \mathcal{C}_{12} & 0 & 0 \\
\mathcal{C}_{21} & \mathcal{C}_{22} & 0 & 0 \\
0 & 0 & 0 & 0 \\
0 & 0 & 0 & 0%
\end{array}%
\right) .  \label{4.7}
\end{equation}
The $\{\theta ,\phi \}$ block of the SCGT therefore reproduces the CFIM
associated with the POVM exactly. This identification provides a nontrivial
consistency check between the two independent constructions, namely the
probability derivatives of Sec.~III and the outcome responses used here. It
reflects the fact that the only parameters accessible to the classical
statistics of the outcomes are the angles, and that the phases, which leave
every $p_{\omega }$ invariant, do not enter this block at all.

\subsubsection{The geometric remainder $\mathcal{I}$}

The remaining part $\mathcal{I}\left( \left\vert \psi _{\mathbf{\lambda }%
}\right\rangle \text{,}\mathrm{E}\left( \varepsilon \right) \right) $ is
nonzero and lives in the phase sector. It assigns the phases ($\mu $ and
$\nu $) a finite, positive semidefinite metric block. This happens even
though the outcome probabilities are completely independent of them.
Physically, this block is generated by the phase responses ($\chi _{\omega
,3}$ and $\chi _{\omega ,4}$) of the individual outcomes. These are
geometric information that classical statistics cannot resolve, but the SCGT
records. It is the quantum part of the semiclassical metric,
\begin{equation}
\mathcal{I}\left( \left\vert \psi _{\mathbf{\lambda }}\right\rangle
\text{,}\mathrm{E}\left( \varepsilon \right) \right) =\left(
\begin{array}{cccc}
0 & 0 & 0 & 0 \\
0 & 0 & 0 & 0 \\
0 & 0 & \mathcal{C}_{33} & \mathcal{C}_{34} \\
0 & 0 & \mathcal{C}_{43} & \mathcal{C}_{44}%
\end{array}%
\right) ,  \label{4.8}
\end{equation}
which is the geometry of the state along the phases as transmitted through
the measurement.

\subsection{Imaginary part of the SCGT}

The imaginary part of $\mathcal{C}\left( \left\vert \psi _{\mathbf{\lambda }%
}\right\rangle \text{,}\mathrm{E}\left( \varepsilon \right) \right) $ can be
written as
\begin{equation}
\text{Im}[\mathcal{C}\left( \left\vert \psi _{\mathbf{\lambda }%
}\right\rangle \text{,}\mathrm{E}\left( \varepsilon \right) \right) ]=%
\mathcal{D}\left( \left\vert \psi _{\mathbf{\lambda }}\right\rangle
\text{,}\mathrm{E}\left( \varepsilon \right) \right) ,  \label{4.9}
\end{equation}
The imaginary part of $\mathcal{C}$ is therefore a real antisymmetric
matrix, which is the semiclassical analogue of the Berry curvature of the
QGT. It is nonzero only in the mixed angle--phase sectors. Its entries, such
as $[\mathcal{D}]_{13}=$Im$\mathcal{C}_{13}$, measure the geometric
(Berry-type) content of the family that is revealed by the measurement.
Thus, we have
\begin{align}
\mathcal{D}\left( \left\vert \psi _{\mathbf{\lambda }}\right\rangle
\text{,}\mathrm{E}\left( \varepsilon \right) \right) &=\left(
\begin{array}{cccc}
0 & 0 & -i\mathcal{C}_{13} & -i\mathcal{C}_{14} \\
0 & 0 & -i\mathcal{C}_{23} & -i\mathcal{C}_{24} \\
-i\mathcal{C}_{31} & -i\mathcal{C}_{32} & 0 & 0 \\
-i\mathcal{C}_{41} & -i\mathcal{C}_{42} & 0 & 0%
\end{array}%
\right) .  \label{4.10}
\end{align}
At $\varepsilon =1$, it reduces to the curvature $\mathcal{G}$ of the QGT,
i.e., $\mathcal{D}\left( \varepsilon =1\right) =\mathcal{G}$.

In summary, the SCGT of the FPSQS is a Hermitian, positive semidefinite
matrix. It decomposes as $\mathcal{C}=\mathcal{F}_{C}+\mathcal{I}+i\mathcal{D
}$. The term $\mathcal{F}_{C}$ is the CFIM of the POVM, located in the
$\{\theta ,\phi \}$ block. The term $\mathcal{I}$ is a
measurement-transmitted quantum metric, located in the phase sector. The
term $\mathcal{D}$ is a semiclassical Berry curvature, located in the mixed
sectors. The SCGT interpolates between two limits. At $\varepsilon =0$, it
reaches the trivial limit, where no information is gathered. At
$\varepsilon =1$, it reproduces the full QGT, where a sharp projective
measurement recovers the pure-state geometry exactly.

\section{The curvature gap}

The extent to which the semiclassical curvature $\mathcal{D}$ departs from
the quantum curvature $\mathcal{G}$ is measured by the difference
$\Delta =\mathcal{G}-\mathcal{D}$. To quantify how much Berry curvature is
lost when the parameters are read out through the POVM $\mathrm{E}\left(
\varepsilon \right) $, we consider the difference
\begin{equation}
\Delta \left( \left\vert \psi _{\mathbf{\lambda }}\right\rangle
\text{,}\mathrm{E}\left( \varepsilon \right) \right) =\mathcal{G}\left(
\left\vert \psi _{\mathbf{\lambda }}\right\rangle \right) -\mathcal{D}\left(
\left\vert \psi _{\mathbf{\lambda }}\right\rangle \text{,}\mathrm{E}\left(
\varepsilon \right) \right) .  \label{5.1}
\end{equation}

Since $\mathcal{D}\left( \varepsilon =1\right) =\mathcal{G}$ and
$\mathcal{D}\left( \varepsilon =0\right) =0$, the difference satisfies
$\Delta \left( \varepsilon =1\right) =0$ and $\Delta \left( \varepsilon
=0\right) =\mathcal{G}$. A sharp projective measurement therefore captures
the full Berry curvature of the state. An uninformative measurement, by
contrast, retains none of it. For intermediate $\varepsilon $, $\Delta $
measures the curvature content that the measurement fails to transmit.
Structurally, $\Delta $ is a real antisymmetric matrix. Its nonvanishing
independent components are $\Delta _{13}=[\mathcal{G}]_{13}-[\mathcal{D}]%
_{13}$, $\Delta _{14}=[\mathcal{G}]_{14}-[\mathcal{D}]_{14}$, $\Delta
_{23}=[\mathcal{G}]_{23}-[\mathcal{D}]_{23}$, and $\Delta
_{24}=-[\mathcal{D}]_{24}$, together with their antisymmetric partners. Note
in particular that $[\mathcal{G}]_{24}=0$ while $[\mathcal{D}]_{24}=$Im$%
\mathcal{C}_{24}\neq 0$ generically. Thus, the semiclassical readout carries
a curvature component in the $\phi $--$\nu $ plane. This component has no
counterpart in the QGT of the pure state. The effect is induced entirely by
the measurement backaction.

\section{Closeness between CFIM and QFIM}

In this section, we analyze the closeness between the CFIM $\mathcal{F}_{C}$
(associated with $\mathrm{E}\left( \varepsilon \right) $) and the QFIM
$\mathcal{F}_{Q}$ of $\left\vert \psi _{\mathbf{\lambda }}\right\rangle $.
For the FPSQS, both CFIM and QFIM are block structured. $\mathcal{F}_{Q}$ is
the real part of the QGT in Eq.~\eqref{3.8}. Its angle block is $\mathcal{F}%
_{Q}^{\theta \phi }=\mathrm{diag}\left( 4,4s_{\theta }^{2}\right) $, and its
phase block $\mathcal{F}_{Q}^{\mu \nu }$ is the lower-right $2\times 2$
block of that equation. $\mathcal{F}_{C}$, in contrast, occupies only the
$\{\theta ,\phi \}$ block. This block $\mathcal{F}_{C}^{\theta \phi }$ was
obtained in Secs.~III and V. In the phase sector, $\mathcal{F}_{C}$ vanishes.

First, in the angle sector, the CFIM approaches the QFIM as the measurement
becomes sharper. At $\varepsilon =1$, the elements computed in Sec.~V reduce
to $\mathcal{C}_{11}=4=\left( \mathcal{F}_{Q}\right) _{11}$, $\mathcal{C}%
_{22}=4s_{\theta }^{2}=\left( \mathcal{F}_{Q}\right) _{22}$, and
$\mathcal{C}_{12}=0=\left( \mathcal{F}_{Q}\right) _{12}$. Hence,
$\mathcal{F}_{C}^{\theta \phi }=\mathcal{F}_{Q}^{\theta \phi }$ for
$\varepsilon =1$. The sharp projective measurement is therefore
informationally optimal for the angles. The classical Cram\'{e}r--Rao bound
saturates the quantum one for $\theta $ and $\phi $. For $0<\varepsilon <1$,
the elements of $\mathcal{F}_{C}^{\theta \phi }$ are suppressed below their
$\varepsilon =1$ values. Indeed, the general bound $\mathcal{F}_{C}\leq
\mathcal{F}_{Q}$ guarantees $\mathcal{C}_{11}\leq 4$ and $\mathcal{C}%
_{22}\leq 4s_{\theta }^{2}$. Consequently, the classical bound becomes
strictly worse than the quantum one.

Second, in the phase sector, the CFIM vanishes identically for every
$\varepsilon $. No outcome probability depends on $\mu $ or $\nu $. Hence,
the phases can never be estimated classically with this measurement, however
sharp it is. Consequently, we have the strict inequality $\mathcal{F}%
_{C}\leq \mathcal{F}_{Q}$ for all $\varepsilon $. The QFIM generically has
full rank $4$. The CFIM, by contrast, has rank at most $2$. Moreover, these
two matrices never coincide.

A natural quantitative measure of their closeness is the squared
Hilbert--Schmidt distance
\begin{equation}
\Vert \mathcal{F}_{Q}-\mathcal{F}_{C}\Vert _{\mathrm{HS}}^{2}=\Vert \mathcal{%
F}_{Q}^{\theta \phi }-\mathcal{F}_{C}^{\theta \phi }\Vert ^{2}+\Vert
\mathcal{F}_{Q}^{\mu \nu }\Vert ^{2},  \label{6.1}
\end{equation}
The angle-sector term vanishes at $\varepsilon =1$ and grows as $\varepsilon
$ decreases. The phase-sector term $\Vert \mathcal{F}_{Q}^{\mu \nu }\Vert
^{2}=[\mathcal{F}_{Q}]_{33}^{2}+2[\mathcal{F}_{Q}]_{34}^{2}+[\mathcal{F}%
_{Q}]_{44}^{2}$ is independent of $\varepsilon $. It cannot be reduced by
any population measurement. This reflects the fundamental blindness of the
POVM to the relative phases. Such blindness can be lifted only by coherent
(off-diagonal) measurements, as noted in Sec.~III. In the same spirit, the
imaginary parts of the two tensors are compared by the difference
$\Delta =\mathcal{G}-\mathcal{D}$, analyzed in the previous section. This
difference vanishes at $\varepsilon =1$ and approaches $\mathcal{G}$ as
$\varepsilon \rightarrow 0$.

\section{Conclusion}

To summarize, we have carried out an explicit information-geometric study of
a class of pure four-parameter single-qutrit states. These states are probed
by a one-parameter family of measurements. The family interpolates between
an uninformative POVM and a sharp projective measurement. We derived the
CFIM. We showed that the phase parameters are invisible to classical
estimation for every measurement in the family, because the outcome
probabilities depend only on the two angles. We then constructed the QGT.
Its real part is the Fubini--Study metric (equivalently, the QFIM), and its
imaginary part is the Berry curvature. We built the SCGT
$\mathcal{C}=\mathcal{F}_{C}+\mathcal{I}+i\mathcal{D}$. It reproduces the
QGT exactly in the projective limit and vanishes in the trivial limit. The
real part of the SCGT splits into the CFIM and a nonnegative
measurement-transmitted metric. The latter carries the phase-sector
geometry. The imaginary part of the SCGT provides a semiclassical Berry
curvature. The difference $\Delta =\mathcal{G}-\mathcal{D}$ quantifies the
curvature content lost by the measurement. Comparing $\mathcal{F}_{C}$ with
$\mathcal{F}_{Q}$, we found that the classical information saturates the
quantum bound in the angle sector at the projective limit. In the phase
sector, however, it can never do so. This is in accordance with the general
obstruction identified in the SCGT framework \cite{17}.

The FPSQS platform provides an exactly solvable test bed for identifying
optimal measurements that saturate the quantum Cram\'{e}r--Rao bound, for
relating the SCGT to the Holevo bound, and for testing the hierarchy of
saturation conditions of multiparameter quantum metrology \cite{16}.
Extending the present analysis to mixed states (which arise naturally under
decoherence or in the presence of noisy devices) would connect the
semiclassical framework with the Bures--Uhlmann geometry \cite{18}, and the
semiclassical Berry phase associated with the SCGT merits further study, in
particular its topological content for generic POVMs.

\begin{acknowledgments}
This work was supported by the National Natural Science Foundation of China
(Grant No. 12465004).
\end{acknowledgments}

\end{document}